\documentclass[utf8,sort&compress]{frontiersFPHY} 

\usepackage{url,hyperref,lineno,microtype,subcaption}
\usepackage{authblk}
\usepackage{booktabs}
\usepackage{color}
\usepackage{amsmath,amssymb,amsfonts}
\usepackage{mathtools}
\usepackage{xcolor,colortbl}
\usepackage{siunitx}
\usepackage{graphicx}
\usepackage{multicol}
\usepackage{footmisc}
\usepackage[square]{natbib}
\usepackage[onehalfspacing]{setspace}

\def\keyFont{\fontsize{8}{11}\helveticabold }
\def\firstAuthorLast{Scarsoglio {et~al.}} 
\def\Authors{Stefania Scarsoglio\,$^{1,*}$, and Luca Ridolfi\,$^{2}$}
\def\Address{$^{1}$Department of Mechanical and Aerospace Engineering, Politecnico di Torino, Torino, Italy \\
$^{2}$Department of Environmental, Land and Infrastructure Engineering, Politecnico di Torino, Torino, Italy  }
\def\corrAuthor{Stefania Scarsoglio}

\def\corrEmail{stefania.scarsoglio@polito.it}

\begin{document}
\onecolumn
\firstpage{1}

\title[HRV in the deep cerebral and central hemodynamics]{Different impact of heart rate variability in the deep cerebral and central hemodynamics at rest: an \emph{in silico} investigation}

\author[\firstAuthorLast ]{\Authors} 
\address{} 
\correspondance{} 

\extraAuth{}

\maketitle

\begin{abstract}
\textbf{Background:} Heart rate variability (HRV), defined as the variability between consecutive heartbeats, is a surrogate measure of cardiac vagal tone. It is widely accepted that a decreased HRV is associated to several risk factors and cardiovascular diseases. However, a possible association between HRV and altered cerebral hemodynamics is still debated, suffering from HRV short-term measures and the paucity of high-resolution deep cerebral data. We propose a computational approach to evaluate the deep cerebral and central hemodynamics subject to physiological alterations of HRV in an ideal young healthy patient at rest.

\textbf{Methods:} The cardiovascular-cerebral model is composed by electrical components able to reproduce the response of the different cardiovascular regions and their features. The model was validated over more than thirty studies and recently exploited to understand the hemodynamic mechanisms between cardiac arrythmia and cognitive deficit. Three configurations (baseline, increased HRV, and decreased HRV) are built based on the standard deviation (SDNN) of RR beats. For each configuration, 5000 RR beats are simulated to investigate the occurrence of extreme values, alteration of the regular hemodynamics pattern, and variation of mean perfusion/pressure levels.

\textbf{Results:} In the cerebral circulation, our results show that HRV has overall a stronger impact on pressure than flow rate mean values but similarly alters pressure and flow rate in terms of extreme events. By comparing reduced and increased HRV, this latter induces a higher probability of altered mean and extreme values, and is therefore more detrimental at distal cerebral level. On the contrary, at central level a decreased HRV induces a higher cardiac effort without improving the mechano-contractile performance, thus overall reducing the heart efficiency.

\textbf{Conclusions:} Present results suggest that: (i) the increase of HRV \emph{per se} does not seem to be sufficient to trigger a better cerebral hemodynamic response; (ii) by accounting for both central and cerebral circulations, the optimal HRV configuration is found at baseline. Given the relation inversely linking HRV and HR, the presence of this optimal condition can contribute to explain why the mean HR of the general population settles around the baseline value (70 bpm).

\tiny
 \keyFont{ \section{Keywords:} heart rate variability, cardiovascular modeling, cerebral circulation, computational hemodynamics, time-series analysis} 
\end{abstract}

\section{Introduction}


Heart rate variability (HRV), defined as the variability between successive RR heartbeats, has grown as \emph{hot topic}, with a simple "Heart rate variability" topic search currently listing more than 44,000 results on Web of Science. The great interest elicited within the scientific community ranges from psychophysiology \citep{Laborde2017} to exercise \citep{Michael2017}, cardiovascular risk factors \citep{Tsuji1996}, and chronic fatigue syndrome \citep{Meeus2013}. This increasing and wide interest is also due to the fact that HRV measurements are easy to perform, quite reliable and non-invasive. In particular, HRV can be considered a surrogate measure of the cardiac vagal tone \citep{Laborde2017}, although there are controversial aspects in using HRV to determine cardiac vagal tone indep endently from other factors, such as age, physical condition, and the presence of cardiac pathologies \citep{Boyett}. In this paper, we refer to the parasympathetic activity within cardiac regulation as cardiac vagal tone. HRV parameters in time and frequency domain able to reflect cardiac vagal tone \citep{Laborde2017,Malik,Berntson} show that an increased cardiac vagal tone  is associated to a higher HRV, while a reduction of the cardiac vagal tone response is linked to a HRV reduction \citep{Hayano1991,Singh}. Moreover, it has recently being recognized that HRV (as measured by means of different metrics, such as SDNN, RMSSD, pvRSA, and HF), inversely correlates with the heart rate (HR), and this relation should be taken into account when dealing with increased/reduced HRV \citep{Monfredi2014,Shaffer2017,Geus2019,Boyett}. The inverse relation between HR and HRV can be explained considering that a faster HR reduces the interval between successive beats and the opportunity for the RR beating to vary, lowering the HRV. On the contrary, a slower HR increases the cardiac interval and enhances the chance for RR to vary, raising HRV \citep{Shaffer2017}.

From an overall cardiovascular point of view, a HRV increase in the physiological range seems beneficial at rest, while a reduced HRV is symptomatic of a stress action, can be predictor of cardiovascular risk factors and is correlated to higher morbidity and mortality \citep{Tsuji1996,Dekker,Meeus2013}. Thus, the clinical interest has been mainly focused on the prognostic significance of HRV related to the risk factors of cardiovascular pathologies. There is substantial evidence that a decreased HRV enhances several risk factors and is associated to cardiovascular diseases, such as left ventricle hypertrophy \citep{Acharya}, sudden cardiac death \citep{Sessa}, and myocardial infarction \citep{Buccelletti}. On the contrary, lower risk factor profiles are associated with increased HRV \citep{Thayer}. The HRV scenario is opposite under physical effort: for increasing moderate-to-vigorous exercise intensity HRV decreases \citep{Tulppo,Michael2017}. During exercise a HRV decrease is beneficial and forced by an increased HR, however there is an immediate post-exercise recovery of HRV \citep{Michael2017} and exercise training can increase resting HRV \citep{Billman}.

Although the overall HRV impact on the central hemodynamics is quite clear, especially related to cardiac disease conditions, its influence on the cerebral hemodynamics is barely known so far. In fact, definitive association between HRV and cerebral circulation is still missing and mainly involve pathologic conditions, such as chronic fatigue syndrome \citep{Meeus2013,Boissoneault} and stroke \citep{Fyfe}. Attention has recently grown about the possible association between increased HRV and improved cognitive performance. Results are conflicting, as some studies have demonstrated the association \citep{Kim,Shah,Schaich2020}, while some others observed that a reduced HRV does not contribute to cognitive impairment or even dementia \cite{Allan,Britton,Mahinrad,Hazzouri}. To understand this controversy it should be kept in mind that, among many influencing factors such as the sample and the objectives of the study, most of the literature is based on short term measures, which have lower prognostic value than 24-hours HRV \citep{Malik,Nunan,Shaffer2017}.

\noindent Though a possible association is widely debated, the underlying hemodynamic mechanisms are so far only hypothesized and mostly unclear \citep{Schaich2020}. It is presently unknown whether HRV is able to influence the normal pressure and flow rate pattern at distal-capillary level, to alter the mean perfusion and pressure levels, and to enhance the occurrence of extreme values. Detailed information are lacking since current non-invasive techniques, such as transcranial doppler ultrasonography, are not able to offer high-resolution data in terms of pressure and flow rate beyond the circle of Willis. Therefore, in the deep cerebral circulation hemodynamic measurements are unreliable or almost absent. Thus, it can be of great significance understanding and predicting from a computational point of view how deep cerebral hemodynamics is affected by HRV at rest. Despite the intrinsic limits, computational hemodynamics and cardiovascular modeling are becoming increasingly important to isolate specific mechanisms and investigate processes where clinical data are not yet feasible, accurate, or easily measurable.

We propose an \emph{in-silico} study where, through a validated combined cardiovascular-cerebral model, we evaluated the deep cerebral hemodynamics subject to physiological alterations of HRV in an ideal young healthy patient at rest, thereby comparing the cerebral to central hemodynamics responses to HRV changes. With deep cerebral circulation we refer to the microcirculation beyond the circle of Willis (i.e., distal and capillary-venous circulation), while the set of parameters defined through cardiac variables and central arterial pressure characterizes the central hemodynamics. The complete lumped-parameter model is composed by a suitable combination of electrical counterparts (resistances, compliances/elastances, inertances), accounting for the arterial and venous circuits of both systemic and pulmonary circulations, an active representation of the four cardiac chambers, an accurate valve motion description, and a short-term baroreceptor mechanism. The cerebral circulation is divided into three main regions: large arteries, distal arterial circulation, and capillary/venous circulation. Cerebrovascular control mechanisms of autoregulation and $CO_2$ reactivity are taken into account. The present computational approach has been validated and recently exploited in a wide area of applications, such as the hemodynamic response to exercise in atrial fibrillation \citep{Anselmino2017}, the linking mechanisms between cardiac arrythmia and cognitive impairment \citep{Anselmino2016,Scarsoglio2017,Chaos2017,Saglietto2019}, and the cardiovascular deconditioning emerging in altered gravity conditions \citep{Gallo2020}.

\noindent We considered three HRV configurations (baseline, increased HRV, decreased HRV), each of them analyzed over 5000 RR beats, so that our results are not affected by transient behavior and are statistically significant and stable. HRV variations were assessed through SDNN, which is the standard deviation of all RR beats and allowed us to define in a straightforward and univocal way the RR stochastic beating extraction. The model is able to provide the whole pressure and flow rate time-series from proximal to distal circulation as well as beat-to-beat values. The focus was on the possible occurrence of extreme values, such as hypertensive events, alteration of the regular hemodynamics pattern, and variation of mean perfusion/pressure levels. All these aspects are to date mostly unexplored, but extremely useful to understand the role of HRV on the cerebral circulation and how, in turn, the hemodynamic alteration can impact the cognitive sphere. The present work can offer precious insights by: (i) isolating the net basic mechanisms induced by HRV changes into the cerebral circulation in healthy resting conditions, and (ii) quantifying how differently HRV acts on the cerebral hemodynamics with respect to the central hemodynamics, thus fostering necessary future clinical measurements within this topic.

\section{Materials and Methods}

\subsection{HRV configurations}
We considered three HRV configurations: baseline condition (reported in the following sections with blue color), increased HRV (in red color) and decreased HRV (in black color). To assess HRV variations we adopted the simplest HRV metric, i.e., SDNN - which is the standard deviation of all RR beats. SDNN belongs to the time-domain analyses, which are computationally simpler and easier to apply than frequency-domain analyses \citep{Michael2017}. Since we focused on resting configurations lasting hours, we chose SDNN because it is the gold standard for long-term measurements \citep{Malik,Shaffer2017}. Moreover, the SDNN adoption allows us to define in a straightforward and univocal way the RR beating extraction. In terms of SDNN, HRV was found to be dependent on the circadian day/night cycle \citep{Shaffer2017} - e.g., 93.06 [ms] day, 121.31 [ms] night, 101.71 [ms] 24-hours \citep{Talib} - and to decrease with age in adulthood \citep{vandenBerg}. In general, SDNN recording was lower in resting supine condition (65 [ms] \citep{Lehavi}; 49 [ms] \citep{Nunan}) than in non-resting or upright position (e.g., 93.06 [ms] \citep{Talib}; 127 [ms] \citep{Genovesi}).
It was recently observed that HRV inversely  correlates with HR through an exponential function \citep{Monfredi2014,Kazmi,Shaffer2017,vandenBerg,Geus2019}, and this relation should be taken into account when dealing with increased/reduced HRV.

To consider SDNN for a healthy young person in resting supine condition during day (awake) and account for the interplay between SDNN e HR, we adopted the exponential relation obtained by de Geus et al. \citep{Geus2019} in leisure baseline condition:

\begin{equation}
\textmd{SDNN} = 309.4 \textmd{e}^{-0.021 \textmd{HR}}
\label{Geus}
\end{equation}

The baseline configuration was set at HR=70 bpm and SDNN=71.14 [ms] was obtained from the above relation. Recalling that the beating period is RR=60/HR [s], we define the coefficient of variation as $c_v$=SDNN/RR. The resulting $c_v$ is 0.08 which well represents normal sinus rhythm daily values in resting supine condition (e.g., $c_v$=0.05 \citep{Pikkujamsa}; $c_v$=0.07 \citep{Lehavi}; $c_v$=0.06 \citep{Nussinovitch}). Moreover, $c_v$=0.08 as baseline value is close to the one adopted by our group ($c_v$=0.07) to simulate resting supine sinus rhythm \citep{Scarsoglio2014}. Increased and decreased HRV configurations were achieved by changing  SDNN by +20\% and -20\%, respectively. This threshold represents a significant variation within a range of physiological values. The corresponding HRs were individuated again using Eq. (\ref{Geus}) as proposed by \citep{Geus2019}. Fig. 1a reports the SDNN(HR) curve in resting condition introduced by \citep{Geus2019} with the values here chosen and below a summarizing table of the three cases.

For each configuration, we evaluated 5000 RR beats to have statistically significant and stationary results. RR beatings were extracted from an \emph{in silico} pink-correlated Gaussian distribution, which well reproduces the typical beating features of normal sinus rhythm recorded in vivo \citep{Hayano,Pikkujamsa,Hennig,Scarsoglio2014}, having mean and standard deviation values identified as above. In the range SDNN $\in$ [30, 100] ms, we also evaluated the RMSSD (i.e., the root mean square of successive RR interval differences) for the RR series extracted with the adopted time-correlation features and composed by 10$^8$ beats. A high linear correlation value ($R^2=0.99$) was found between RMSSD and SDNN, with fitting law RMSSD=0.65 SDNN + 0.54, as reported in Fig. 1b. Although RMSSD and SDNN don't have the same physiological origin and only RMSSD can be assumed to fully reflect cardiac vagal activity \citep{Laborde2017,Malik,Berntson}, SDNN is a reliable proxy of RMSSD for the considered RR time series. Bottom panels of Fig. 1 display the RR series (panel c) and the corresponding probability density functions (PDFs) for the three HRV cases here investigated (panel d).

\subsection{Cardiovascular and cerebral circulation modeling}

Following RR extraction, the cardiovascular-cerebral model was run to obtain the hemodynamic cerebral signals. The complete lumped model is composed by electrical components able to reproduce the response of the different (cardiac and vascular) regions and their features, in terms of resistance $R$ (diffusive effects), inertance $L$ (inertial effects), and compliance $C$ or elastance $E$ (elasticity/contractility effects). The cardiovascular dynamics includes the arterial and venous circuits of both systemic and pulmonary circulations, an active representation of the four cardiac chambers, and an accurate valve motion description. A short-term baroreceptor mechanism is also modeled, accounting for the inotropic effect of both ventricles, as well as the control of the systemic vasculature (peripheral arterial resistances, unstressed volume of the venous system, and venous compliance). The chronotropic effects due to the heart rate regulation are intrinsically taken into account by the RR extraction. The cerebral circulation is divided into three principal regions: large arteries, distal arterial circulation, and capillary/venous circulation. Cerebrovascular control mechanisms of autoregulation and $CO_2$ reactivity are taken into account. The model is expressed in terms of pressures, $P$, flow rates, $Q$, volumes, $V$, and valve opening angles, $\theta$.

\noindent The cardiovascular model has been proposed and validated over more than thirty studies to check the consistency of the hemodynamic response during AF: extensive details of this evaluation are reported in \citep{Scarsoglio2014}. Then, the model has been exploited to study the impact of atrial fibrillation on the cardiovascular system \citep{Anselmino2015,CMBBE2016,PeerJ2016,Anselmino2017}. The complete cardiovascular-cerebral model has been used to understand the hemodynamic mechanisms between atrial fibrillation and cognitive deficit \citep{Anselmino2016,Scarsoglio2017,Chaos2017,Saglietto2019}. A full description of the governing equations and model parameters is given in the Supplementary Material.

We here focused - in terms of pressure $P$ and flow rate $Q$ - on the left internal carotid artery-middle cerebral artery (ICA-MCA) pathway, which already turned out to be representative of the hemodynamics from proximal to distal cerebral districts \citep{Anselmino2016,Scarsoglio2017,Chaos2017,Saglietto2019}. The left ICA-MCA path starts at the internal carotid level ($P_a$: systemic arterial pressure; $Q_{ICA,left}$: left internal carotid flow rate), goes through the middle cerebral artery ($P_{MCA,left}$: left middle cerebral artery pressure; $Q_{MCA,left}$: left middle cerebral artery flow rate), includes the middle distal regions ($P_{dm,left}$: left middle distal pressure; $Q_{dm,left}$: left middle distal flow rate), and ends with capillary-venous districts ($P_c$: cerebral capillary pressure; $Q_{pv}$: proximal venous flow rate). A schematic representation of the systemic-pulmonary circulation and ICA-MCA pathway is reported in Figure 2, together with examples of left ventricle PV loops, aortic flow rate ($Q_{ao}$), systemic arterial pressure ($P_a$), and cerebral capillary pressure ($P_c$) time-series for different HRV configurations.

\subsection{Variable definition and data analysis}

We recall the definition of mechano-energetic and oxygen consumption indexes, which will be used to describe the central hemodynamics. End-systolic left ventricular volume, $V_{lves}$ [ml], is the left ventricle volume at the closure of the aortic valve, while end-diastolic left ventricular volume, $V_{lved}$ [ml], corresponds to the closure of the mitral valve. Stroke volume is defined as $SV=(V_{lved} - V_{lves})$ [ml], ejection fraction is $EF = V/V_{lved} \cdot 100$ [\%]. Cardiac output is $CO = SV \cdot HR$ [l/min], stroke work per minute, $SW/min$ [J/min], is measured as the area within the left ventricle pressure-volume loop per beat. Oxygen consumption is evaluated through three estimates \citep{Westerhof}: (i) the rate pressure product, $RPP=P_{a,syst} \cdot HR$ [mmHg/min], where $P_{a,syst}$ is the aortic systolic pressure; (ii) the tension time index per minute, $TTI/min = \overline{P}_{lv} \cdot RR \cdot HR$ [mmHg s/min], where the symbol $\overline{f}$ indicates the mean value of the generic hemodynamic variable $f$ averaged over a RR beat (in this case $f=P_{lv}$); and (iii) the pressure volume area per minute, $PVA/min = (PE + SW) \cdot HR$, where $PE = [P_{lves} \cdot (V_{lves} - V_{lv,un})/2-P_{lved} \cdot (V_{lved}-V_{lv,un})/4]$ is the elastic potential energy ($V_{lv,un}$ = 5 ml is the unstressed left ventricle volume), $SW$ is the stroke work, $P_{lves}$ and $P_{lved}$ are the end-systolic and end-diastolic left ventricle pressures. The left ventricular efficiency $LVE$ is defined by the ratio $SW/PVA$.

The cerebral hemodynamics is assessed by means of two main approaches: (i) analysis of the continuous time-series, where the signal is continuous and defined by all the temporal instants of the whole time-series, and (ii) beat-to-beat analysis, where the signal was discretized and one-per-beat data were obtained. In so doing, beat-to-beat signals are composed by 5000 values, corresponding to the 5000 RR beats simulated.

\noindent For the analysis of the continuous time-series, extremely high or low cerebral hemodynamic values were evaluated through the percentile analysis of the hemodynamic signals. We adapted the definition used for studying atrial fibrillation \citep{Anselmino2016,Saglietto2019}, to assess the possibility of extreme events related to HRV changes. For the increased and decreased HRV cases, we estimated to which percentile the baseline (10th and 90th) reference thresholds correspond, by quantifying the probability of assuming rare values. Top panels of Fig. 3 (a, b) provide a representative graphical representation of the percentile analysis for $P_{dm,left}$. The 10th and 90th percentiles individuated in baseline correspond to the 2nd and 85th percentiles in the decreased HRV case (panel a), while the 10th and 90th baseline percentiles correspond to the 22nd and 94th percentiles in the increased HRV case (panel b). In so doing, we quantify whether HRV changes are able to make extreme values more frequently reached with respect to the baseline case (e.g., in Fig. 3b the 10th baseline percentile corresponds to the 22nd percentile for increased HRV).

\noindent For the beat-to-beat analysis, the $i-th$ element of the discretized series may contain the average value ($\overline{Q}$ and $\overline{P}$), the maximum ($Q^{max}$ and $P^{max}$) and minimum ($Q^{min}$ and $P^{min}$) values, as well as the pulse value ($Q^{pv}$ and $P^{pv}$, defined as the difference between maximum and minimum values) of the related haemodynamic variables computed over the $i-th$ beat. In Fig. 3d examples of $\overline{P}$, $P^{max}$, $P^{min}$, and $P^{pv}$ are represented for $P_{dm,left}$.

\noindent The average values per beat, $\overline{P}$ and $\overline{Q}$, were still exploited to enrich the beat-to-beat analysis, by accounting for the persistence of extreme values over the whole beat \citep{Anselmino2016,Saglietto2019}, and not only the occurrence of instantaneous peak values. We introduce hypoperfusions (or hypotensive events), which take place when the mean flow rate per beat $\overline{Q}$ (or mean pressure per beat $\overline{P}$) is below the 10th percentile referred to the whole baseline signal. On the contrary, hyperperfusions (or hypertensive events) emerge when $\overline{Q}$ (or $\overline{P}$) is above the 90th percentile of the baseline condition. Note that, by definition, extreme events cannot emerge in the baseline configuration. Bottom panels of Fig. 3 (c, d) show examples of hypertensive and hypotensive events for $P_{dm,left}$.

\section{Results}

\subsection{Continuous $P$ and $Q$ time-series analysis}
Table \ref{complete_statistics} reports the main statistics (mean $\mu$, standard deviation $\sigma$, and coefficient of variation $c_v$) of the continuous time-series of the hemodynamic variables along the ICA-MCA pathway. Figure \ref{Fig_PDF_complete} shows probability density functions (PDFs) of the whole hemodynamic time-series for the three configurations. Considering the continuous hemodynamic signals, HRV influenced much more pressure than flow rate series. In fact, mean values were overall maintained (see Table \ref{complete_statistics}) and PDFs revealed self-similar features (see Fig. \ref{Fig_PDF_complete}) for flow rates $Q$, while HRV significantly affected mean values and PDFs of pressures $P$ (only $P_c$ at the end of the pathway regained a self-similar shape for the three configurations). At a given district, HRV impacted $\sigma$ relative variations  with respect to baseline similarly for both pressure and flow rate, while $c_v$ relative variations induced by HRV with respect to baseline were more evident towards the distal-capillary circulation.

It is then useful to evaluate how extreme values of the continuous time-series $P(t)$ and $Q(t)$ distribute as HRV changes, by considering the percentile analysis (as represented in Fig. 3a-b). By assessing to which percentile the baseline reference thresholds correspond in the configurations with altered HRV, we can quantify whether HRV is able to modify the probability of reaching extremely high/low values (see Fig. \ref{Fig_percentile}). Distal-capillary regions were the most affected by HRV: the 10th percentile for baseline $P_{dm,left}$ corresponds to over the 20th percentile for increased HRV, while the 90th percentile for baseline $P_c$ corresponds to about the 82th percentile for the reduced HRV configuration (see top panels of Fig. \ref{Fig_percentile}). It is worth noting that a similar scenario was found for flow rates as well. The increased HRV enhanced extreme values of the continuous time-series towards the venous circulation: the 10th percentile corresponds to the 17th, while the 90th to the 82th (see bottom panels of Fig. \ref{Fig_percentile}).

\subsection{Beat-to-beat analysis}

The scenario of higher impact of HRV on cerebral pressures than flow rates was also confirmed in the beat-to-beat analysis. Table \ref{beat_to_beat} displays mean and standard deviation values for the minimum ($P^{min}$ and $Q^{min}$), average ($\overline{P}$ and  $\overline{Q}$), maximum ($P^{max}$ and $Q^{max}$), and pulse ($P^{pv}$ and $Q^{pv}$) values per beat of pressures $P$ and flow rates $Q$ as computed over 5000 RR beats. PDFs of $\overline{P}$ and $\overline{Q}$ along the ICA-MCA pathway are depicted in Fig. \ref{Fig_PDF_mean}. The altered HRV did not substantially modify $\overline{Q}$ values, with relative variations of increased and decreased HRV $<$0.1\% with respect to baseline, see the $\overline{Q}$ column in Table \ref{beat_to_beat} and the corresponding PDFs (see Fig. \ref{Fig_PDF_mean}, right panels). Relative variations of minimum ($Q^{min}$) and maximum ($Q^{max}$) values per beat were within 8\%. $\overline{P}$ values as well as the corresponding PDFs were instead much more affected by HRV: only at the capillary level, similar $\overline{P}_c$ values were recovered for the three HRV configurations (with relative variations $<$0.1\%, see bottom left panel of Fig. \ref{Fig_PDF_mean} and the mean value column of Table \ref{beat_to_beat}), while in all the other regions relative $\overline{P}$ variations were within 3\%. Relative variations of maximum ($P^{max}$) and minimum ($P^{min}$) values per beat did not exceed 5\% with respect to the baseline configuration (see Table \ref{beat_to_beat}).

\noindent The different impact of HRV change on pressure and flow rate can be explained noting that cerebral control mechanisms act on the continuous flow rate values at the distal level only, by modifying pial arterial-arteriolar compliances and resistances \citep{Ursino}. Thanks to the autoregulation effects, mean values of the continuous flow rate time-series (Table \ref{complete_statistics}) and beat-to-beat $\overline{Q}$ recordings (Table \ref{beat_to_beat}) were preserved unvaried in the three configurations, while the same was not true for cerebral pressure. Despite cerebral autoregulation mechanisms are concentrated in the distal district solely, their effects involve both upstream and downstream regions. In fact, the mean flow rate is guaranteed as constant for the three configurations throughout the ICA-MCA pathway (i.e., from $Q_{ICA,left}$ to $Q_{pv}$, see $Q$ mean values in Table \ref{complete_statistics} and $\overline{Q}$ in Table \ref{beat_to_beat}).

Based on what emerged so far, we expect a higher occurrence of extreme events in the beat-averaged pressure, $\overline{P}$, than beat-averaged flow rate, $\overline{Q}$. Table \ref{hypo_hyper} reports the total number of rare one-beats hypotensive, hypertensive, hypoperfusion, and hyperperfusion events. We recall that these events, by definition, cannot occur in the baseline configuration, which is taken to set the reference thresholds (10th and 90th percentiles). For the other configurations, one of these events happens at a specific district if the mean value per beat for $\overline{P}$ or $\overline{Q}$ reaches extremely high (above the 90th percentile of the baseline configuration) or low (below the 10th percentile of the baseline configuration) values. In particular, out of 5000 RR beats we counted 42 hypotensive events at distal level for the increased HRV configuration, and 69 hypertensive events for reduced HRV. No hypoperfusions were found, while other more occasional events, such as 9 distal hyperperfusion events, 6 capillary hypertensive events, and 4 distal hypertensive events, were induced by the higher HRV configuration. The reason of the low occurrence of extreme perfusive events can be understood by observing that the widening of the PDFs for the continuous flow rate time-series caused by increased HRV (see the red curves in right panels of Fig. \ref{Fig_PDF_complete}) is related to instantaneous peak values rather than enduring events in time. Thus, increased HRV was not able to trigger significant extreme perfusive events over a whole beat. In fact, PDFs of $\overline{Q}$ values were quite coincident for the three configurations (right panels of Fig. \ref{Fig_PDF_mean}), leading to very sporadic hyperperfusion events and no hypoperfusion events at all. It should be also recalled that HRV variations were taken in a physiological range ($\pm$20\% of the baseline value), therefore it is reasonable not expecting a large number of extreme events based on averaged-beat variables.

\section{Discussion}

The present study aims at computationally investigating the role of HRV on the deep cerebral and central hemodynamics. SDNN has been adopted as a proxy measure for HRV and, although it does not have the same physiological origin reflecting cardiac vagal activity as RMSSD, it is highly correlated with the latter and more suited to the proposed computational approach. HRV seems to have overall a stronger impact on $\overline{P}$ than $\overline{Q}$ values, due to autoregulation effects acting to maintain an adequate average cerebral perfusion. However, in terms of extreme (minimum and maximum) values and percentile distributions, HRV similarly altered both pressure and flow rate. By comparing reduced and increased HRV, this latter induced a higher probability of altered mean values and extreme values (the only exception was the $P_{dm,left}$ case with reduced HRV and hypertensive events), and is therefore more detrimental at distal cerebral level.

As a further confirmation of the worsening linked to a higher HRV, it is observed that starting with an increased HRV, the variability of hemodynamic variables increased from proximal to distal regions along the left ICA-MCA pathway. On the contrary, starting with a reduced HRV, the hemodynamic variability decreased towards the deep cerebral circulation. This behavior is evident for the continuous signals in terms of $c_v$ ratio (as reported in Fig. \ref{Fig_cv}), as well as at beat-to-beat level, through relative variations of pulse values, $P^{pv}$ and $Q^{pv}$ (see Fig. \ref{Fig_ratio_PP}). It should be noted that the relative variations of $P^{pv}$ and $Q^{pv}$ showed this increasing/decreasing trend, although $P^{min}$, $P^{max}$, $Q^{min}$, and $Q^{max}$ did not singularly present, as mentioned before, a similar behavior:
relative variations of $P^{min}$ and $P^{max}$ ($Q^{min}$ and $Q^{max}$) with respect to baseline did not exceed 5\% (8\%) along the ICA-MCA pathway (please refer to Table \ref{beat_to_beat}, columns with minimum and maximum values). Moreover, as displayed by Figures \ref{Fig_cv} and \ref{Fig_ratio_PP}, the behavior was analogously found for both pressure and flow rate.

Even though all results fell within a range of physiological values, increased HRV seems more deleterious than reduced HRV along the proximal-to-distal cerebral pathway. An unsafe increased pulsatility of the hemodynamic signals was found especially at capillary-venous levels, and can promote extremely high/low pressure and flow rate values.

This result can be explained recalling what observed for the cerebral hemodynamics during atrial fibrillation \citep{Scarsoglio2017}. The intrinsic structural latency revealed by the cerebral microcirculation and combined with a sequence of in-series irregular RR beats produced an alteration of the cerebral circulation. Indeed, due to the delay related to mechanical and structural properties, the inertia of the system increased towards the microvasculature.

\noindent Atrial fibrillation represents a succession of disturbances in terms of RR beating, which are then transmitted from the carotid level to the deep cerebral circulation. Each of these perturbations singularly produced an alteration of the cerebral circulation. The continuous sequence of transient perturbations at the carotid entrance represented by the fibrillated beating did not allow the system to recover the physiological state before another disturbance arrives. When this disturbance chain spread throughout the cerebral vessel network, the distal and capillary regions remained altered for longer. The behavior is similar to a system of springs in series and parallel, which is externally forced at one end: the stiffness of each spring combines with the others and the oscillation survives even if the external perturbation is ceased. It is important to note that this mechanism occurred regardless of whether the beat in atrial fibrillation is accelerated or not. In fact, the comparison between NSR and atrial fibrillation \citep{Scarsoglio2017} was proposed at the same mean heart rate, highlighting the role of RR variability in promoting the observed hemodynamic alteration.

\noindent With due proportions, the basic mechanism detected in atrial fibrillation seems to be similar in the case of increased HRV. Clearly, fibrillated outcomes were much more exacerbated and pathological since variability was much higher, and the RR beating was uncorrelated too. However, also here the component of increased variability seems to plausibly combine with the intrinsic latency of the system, causing the observed altered scenario.

The hemodynamic framework is inverted if we focus at central level, in terms of cardiac contractility, efficiency end energetic indexes (see Table \ref{central_statistics}). The decreased HRV configuration showed an increase of oxygen consumption indexes (+16\% $RPP$, +6\% $TTI/min$, +11\% $PVA/min$) and cardiac work (+9\% $EW/min$), in front of a reduction of the ejection fraction (-5\%) and a limited increase of cardiac output (+6\%). The reduction of the left ventricle efficiency $LVE$, though slight (-2\%), confirms that a higher cardiac effort did not translate into a better mechanical and contractile cardiac performance. Moreover, variations of mean central flow rate (in terms of $CO$) did not entail average changes of the flow rate in the deep cerebral circulation (see cerebral mean flow rates in Tables \ref{complete_statistics} and 2), due to the autoregulation mechanisms at the distal cerebral level. In other words, the greater central energy expenditure did not produce any gain in terms of mean flow rate in the cerebral microcirculation.

To extend the discussion, we considered a configuration of heart failure as representative of a very different condition from the healthy state and often present as a co-morbidity in the elderly population. With respect to the baseline heart failure condition, we analyzed two HRV configurations also here, increased (+20 \% SDNN) and decreased (-20 \% SDNN) HRV. Details of the heart failure configurations are provided in the Supplementary Material. For the deep cerebral hemodynamics, the percentile analysis and the PDFs of the continuous hemodynamic time-series in the heart failure case are reported and compared with the healthy case in Fig. \ref{Fig_perc_HF}. Although the values achieved are different from the case healthy (see the blue thick and thin curves), the HRV variations with respect to the baseline are quite close to the healthy case (see the red/black thick and thin curves), especially in the distal regions. Table 5 shows the values of the central hemodynamic parameters in the heart failure case. Compared to the healthy case, the values achieved are very different and show a rather compromised hemodynamic framework (e.g., at baseline $CO$ = 3.55 l/min, $EF$ = 16.41 \%, $LVE$ = 0.31). However, the HRV variations compared to baseline are in agreement with the healthy case (decreased HRV: +12\% $RPP$, +3\% $TTI/min$, +10\% $PVA/min$, +4\% $EW/min$, -8\% $EF$, +3\% $CO$, -5\% $LVE$).

For both healthy and heart failure cases, at central level there is a qualitative agreement with what has been widely observed in literature, that is an increased HRV is beneficial for the cardiac performance and efficiency. At cerebral level, where instead results are still debated, an increase of HRV only - regardless any other variation, compensatory mechanism or pathological condition - seems to be associated to a worse hemodynamic response. A plausible interpretation of these findings with respect to the possible association between HRV and cognitive performance observed in some literature studies is that we only considered the net contribution of HRV alterations, without any complementary and further variation of the sympathetic system. In fact, changes of the sympathetic activity are hardly quantifiable and out of the objectives of the present study. According to the neurovisceral integration model vagally-mediated HRV metrics, such as RMSSD and HF, are expected to be linked to cognitive performance, but not SDNN \citep{Thayer_NVM}. Based on our outcomes the increase of HRV, measured by means of SDNN, does not seem to be sufficient \emph{per se} to trigger a better cerebral hemodynamic response.

\subsection{Limitations}

The present computational approach has some limiting aspects related to the modeling hypotheses. We considered the same young healthy resting subject as forced through three different HRV configurations. In particular, the three configurations only differ by the entrance inputs, the beating sequence RR, while the remaining hemodynamic framework is set as in baseline condition, regardless of other compensatory mechanisms that may occur or altered sympathetic activity. Cerebral autoregulation and autonomic regulation are not fully coupled, but this interplay is present in one-way direction only: through the baroreceptor mechanisms acting on the systemic arterial pressure, autonomic regulation influences the cerebral autoregulation, but there are no feedbacks from cerebral blood flow to autonomic control. Moreover, coronary circulation was not included into the model. In the end, we evaluated a monitoring temporal window of about one hour (5000 RR beats), thus no long-term remodelling effects related to increased or decreased HRV were taken into account.

\section{Conclusions}
In conclusion, by recalling the relation inversely linking SDNN and HR \citep{Monfredi2014,Shaffer2017,Geus2019}, at deep cerebral level a higher SDNN is worse than a higher HR. On the contrary, at central level a higher HR is more negatively impacting than a higher SDNN. From an overall point of view which contemporarily accounts for both central and deep cerebral circulations, present results suggested that the optimal HRV configuration is found at baseline. In fact, an increased HRV (lower HR) is detrimental for the cerebral circulation causing possible hypertensive events and extreme pressure and flow rate values, while a decreased HRV (higher HR) induces a higher cardiac effort without improving the mechano-contractile performance, thus overall reducing the heart efficiency. The presence of this optimal condition is the main finding and can contribute to explain why the mean HR of the general population settles around 70 bpm: the baseline value is a good compromise able to guarantee an adequate hemodynamic response for both central and deep cerebral regions, preserving from extreme cerebral hemodynamic events and at the same time maintaining satisfactory cardiac efficiency.

\section*{Conflict of Interest Statement}

The authors declare that the research was conducted in the absence of any commercial or financial relationships that could be construed as a potential conflict of interest.

\section*{Author Contributions}

All authors conceived and designed the experiments. SS performed the experiments. All authors analyzed the data, contributed reagents/materials/analysis tools. SS wrote the paper. All authors reviewed and approved the final version of the manuscript.


\section*{Data Availability Statement}
The datasets generated for this study are available on request to the corresponding author.

\bibliographystyle{frontiersinHLTH&FPHY} 
\bibliography{test}

\section*{Figure captions}


\begin{figure}[h!]
\begin{center}
\includegraphics[width=0.8\columnwidth]{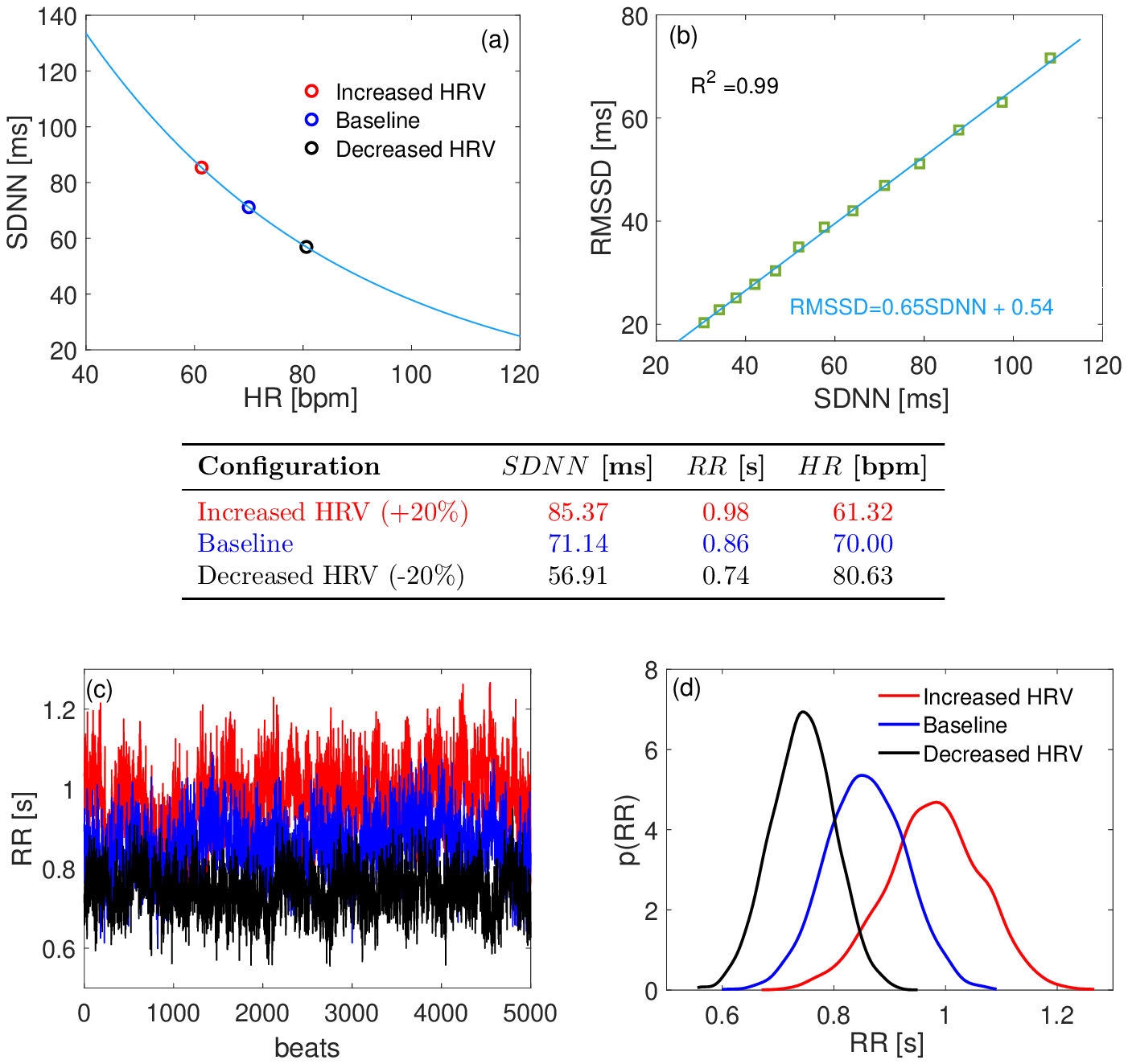}
\end{center}
\caption{Top panels (a)-(b): Relation SDNN(HR) and RMSSD(SDNN). Middle panel: table with the chosen values of the three configurations. Bottom panels (c)-(d): RR series and probability density functions of the three configurations. Red: increased HRV; blue: baseline; black: decreased HRV.}
\label{RR_HR_SDNN}
\end{figure}

\begin{figure}[h!]
\begin{center}
\includegraphics[width=\columnwidth]{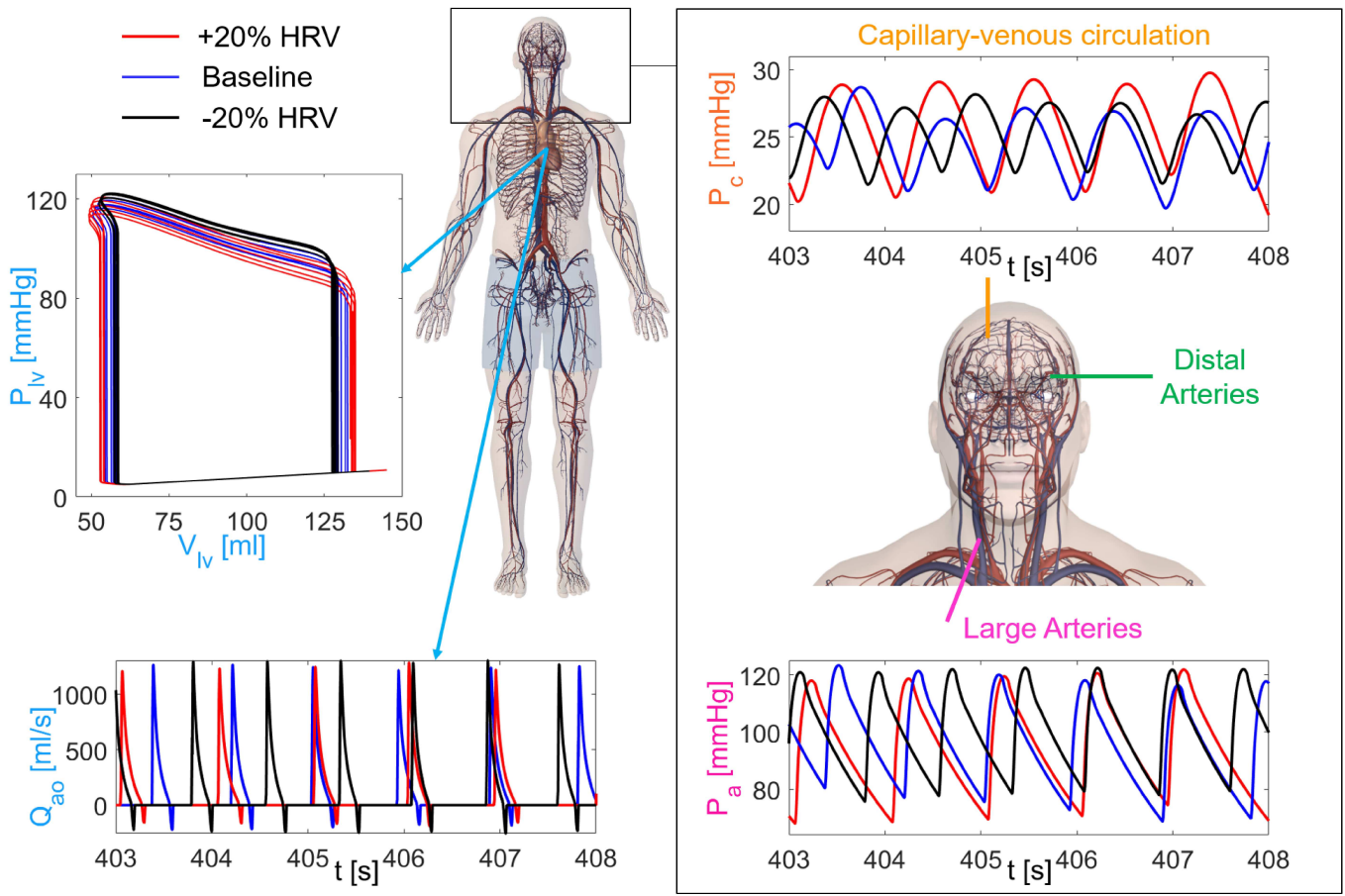}
\end{center}
\caption{Sketch of the cardiovascular-cerebral model for the three HRV configurations (red: increased HRV, blue: baseline, black: decreased HRV). (left) Central and systemic-pulmonary circulation with representative time-series of the aortic flow rate ($Q_{ao}$) and left ventricle PV loops (5 seconds length). (right) Cerebral circulation with the three main regions (large arteries, distal arteries, capillary-venous circulation), together with examples of the systemic arterial pressure, $P_a$ (bottom), and cerebral capillary pressure, $P_c$ (top), time-series.}
\label{Fig2}
\end{figure}

\begin{figure}[h!]
\begin{center}
\includegraphics[width=0.8\columnwidth]{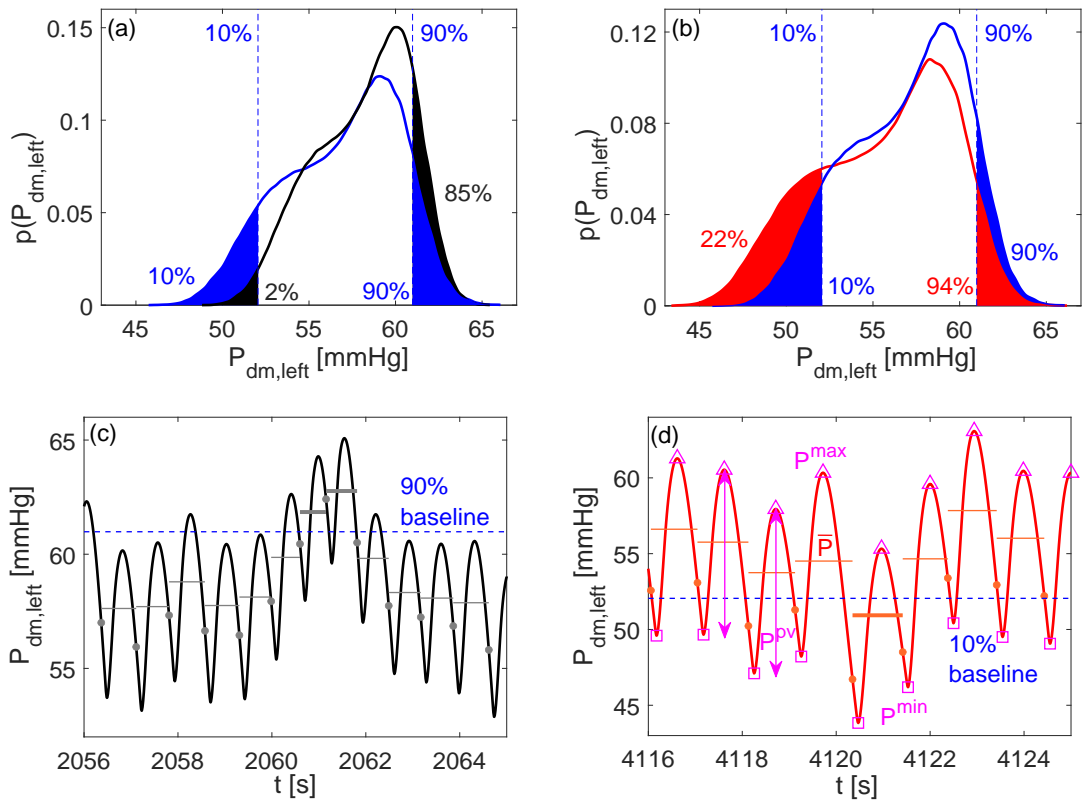}
\end{center}
\caption{(a)-(b) Examples of percentile evaluation for $P_{dm,left}$ ($p(P_{dm,left})$ is the probability density function). 10\% and 90\% blue dashed lines individuate the 10th and 90th percentiles in the baseline configuration (panels a-b, blue areas), while they correspond to the 2nd and 85th percentiles in the decreased HRV configuration (panel a, black areas), and to the 22nd and 94th percentiles in the increased HRV configuration (panel b, red areas). (c)-(d) $P_{dm,left}$ examples of hypertensive event lasting 2 beats in the decreased HRV configuration (average pressure per beat, $\overline{P}$, is represented by grey horizontal lines and hypertensive events are marked with bold grey lines) and hypotensive event in the increased HRV configuration ($\overline{P}$ is represented by red horizontal lines and hypothensive events are marked with bold red lines). The 10th and 90th percentile baseline thresholds are displayed through the dashed blue horizontal lines, while full circles indicate the heartbeat extremes. In panel (d) examples of $P^{max}$ ($\triangle$), $P^{min}$ ($\Box$), and $P^{pv}$ ($\leftrightarrow$) values per beat are evidenced for $P_{dm,left}$.}
\label{Fig3_perc_etreme}
\end{figure}

\begin{figure}[h!]
\begin{center}
\includegraphics[width=0.8\columnwidth]{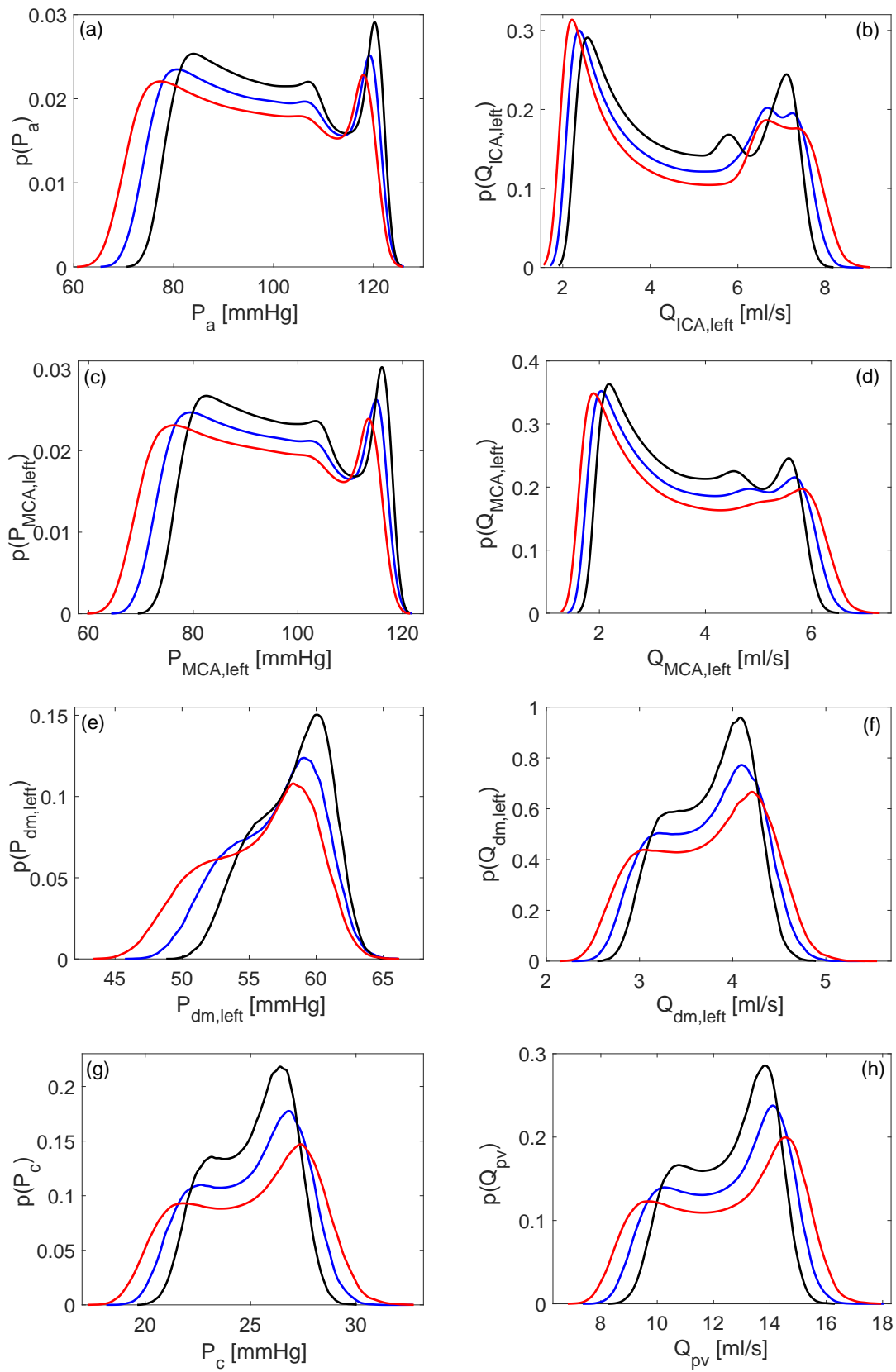}
\end{center}
\caption{Probability density functions of the continuous hemodynamic time-series along the ICA-MCA pathway. Left panels: pressures. Right panels: flow rates. Black: decreased HRV; blue: baseline; red: increased HRV.}
\label{Fig_PDF_complete}
\end{figure}

\begin{figure}[h!]
\begin{center}
\includegraphics[width=0.8\columnwidth]{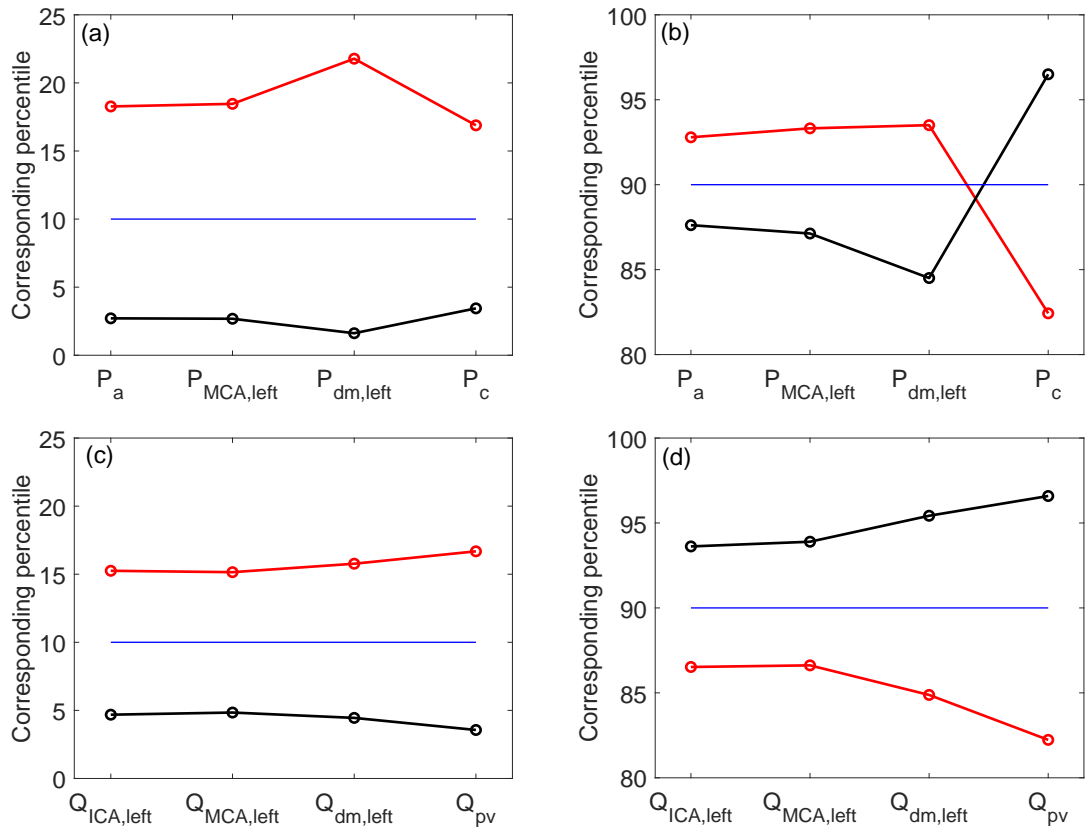}
\end{center}
\caption{Percentile values in increased (red) and decreased (black) HRV to which the 10th (left panels) and 90th (right panels) percentiles in baseline configuration correspond. Pressure (top panels) and flow rate (bottom panels) along the ICA-MCA pathway.}
\label{Fig_percentile}
\end{figure}

\begin{figure}[h!]
\begin{center}
\includegraphics[width=0.8\columnwidth]{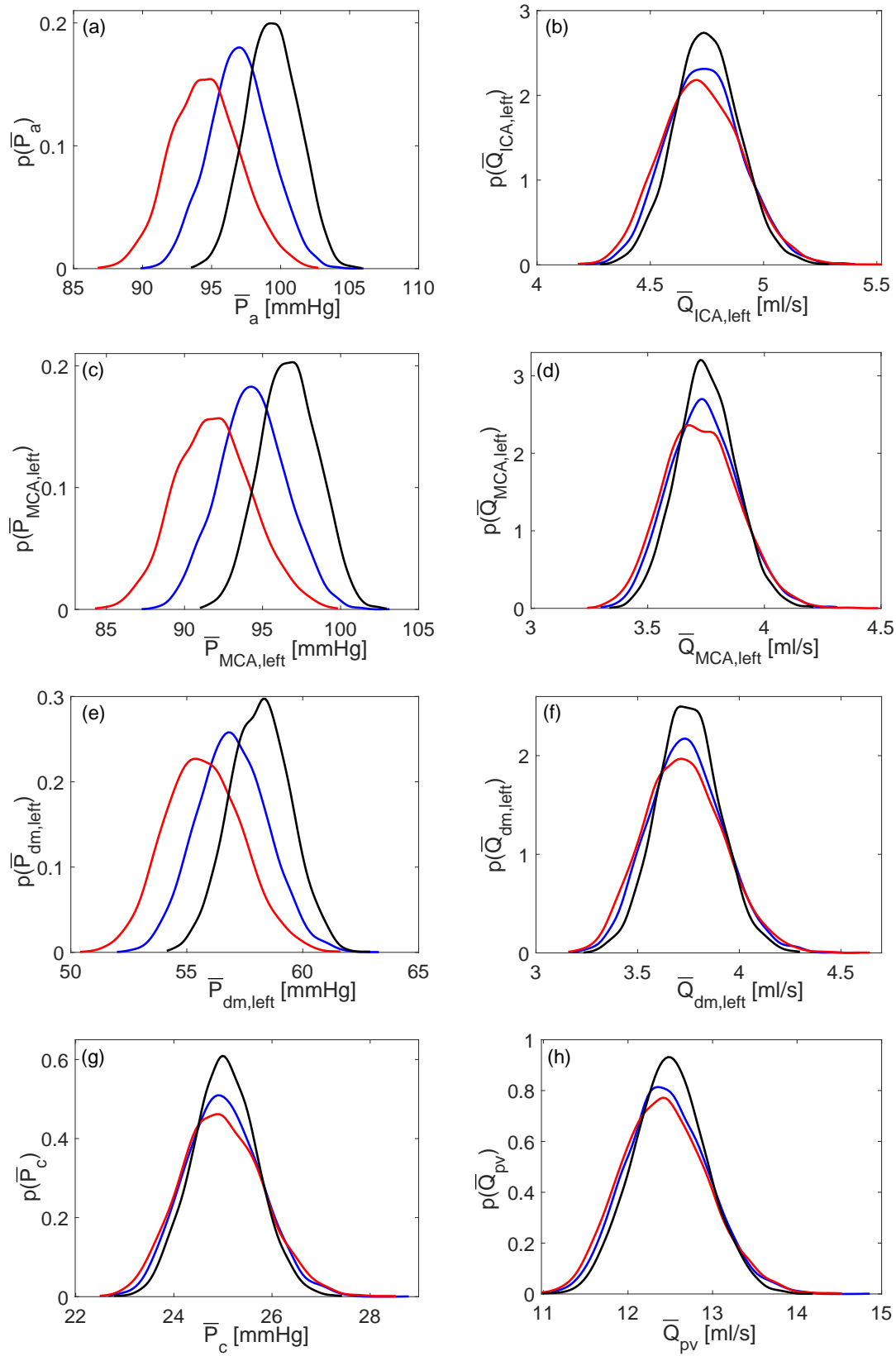}
\end{center}
\caption{Probability density functions of pressure and flow rate values averaged per beat, $\overline{P}$ and $\overline{Q}$, along the ICA-MCA pathway. Left panels: pressures $\overline{P}_a$, $\overline{P}_{MCA,left}$, $\overline{P}_{dm,left}$, $\overline{P}_{c}$. Right panels: flow rates $\overline{Q}_{ICA,left}$, $\overline{Q}_{MCA,left}$, $\overline{Q}_{dm,left}$, $\overline{Q}_{pv}$. Black: decreased HRV; blue: baseline; red: increased HRV.}
\label{Fig_PDF_mean}
\end{figure}

\begin{figure}[h!]
\begin{center}
\includegraphics[width=0.8\columnwidth]{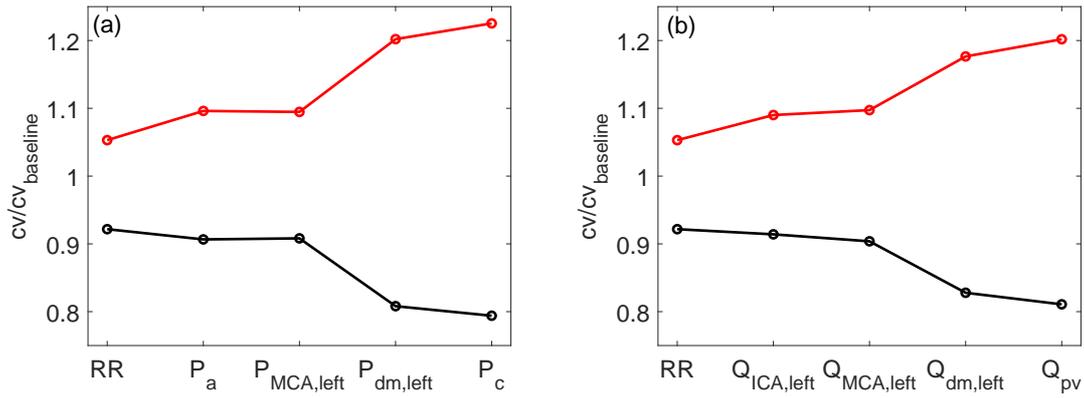}
\end{center}
\caption{Ratio between the coefficient of variation $c_v$ in increased/decreased (red/black) HRV cases and the coefficient of variation in baseline condition. Ratios for RR time-series and pressure/flow rate continuous time-series along the ICA-MCA pathaway are shown.}
\label{Fig_cv}
\end{figure}

\begin{figure}[h!]
\begin{center}
\includegraphics[width=0.8\columnwidth]{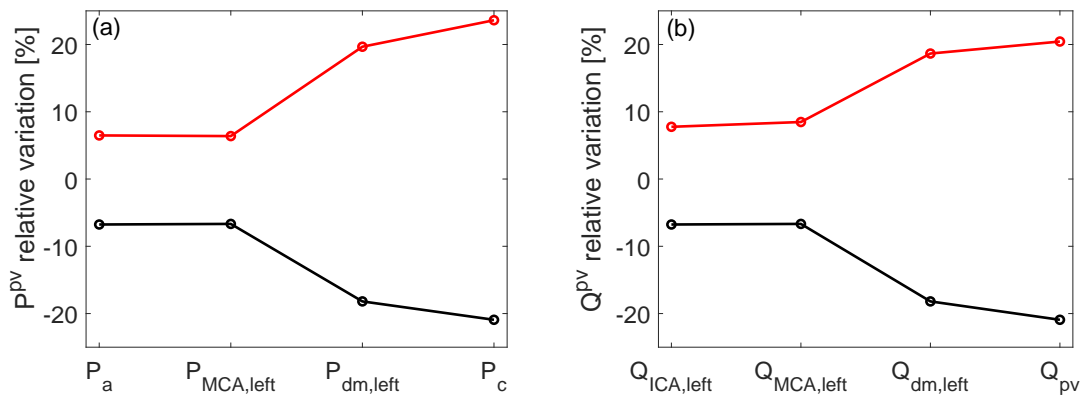}
\end{center}
\caption{Relative variations [\%] of pulse values per beat with respect to baseline: $(P^{pv} – P^{pv}_{baseline})/P^{pv}_{baseline} \cdot 100$ and $(Q^{pv} – Q^{pv}_{baseline})/Q^{pv}_{baseline} \cdot 100$. Pulse values, $P^{pv}$ and $Q^{pv}$, are evaluated at each site of the ICA-MCA pathway. Black: decreased HRV/baseline. Red: increased HRV/baseline.}
\label{Fig_ratio_PP}
\end{figure}

\begin{figure}[h!]
\begin{center}
\includegraphics[width=\columnwidth]{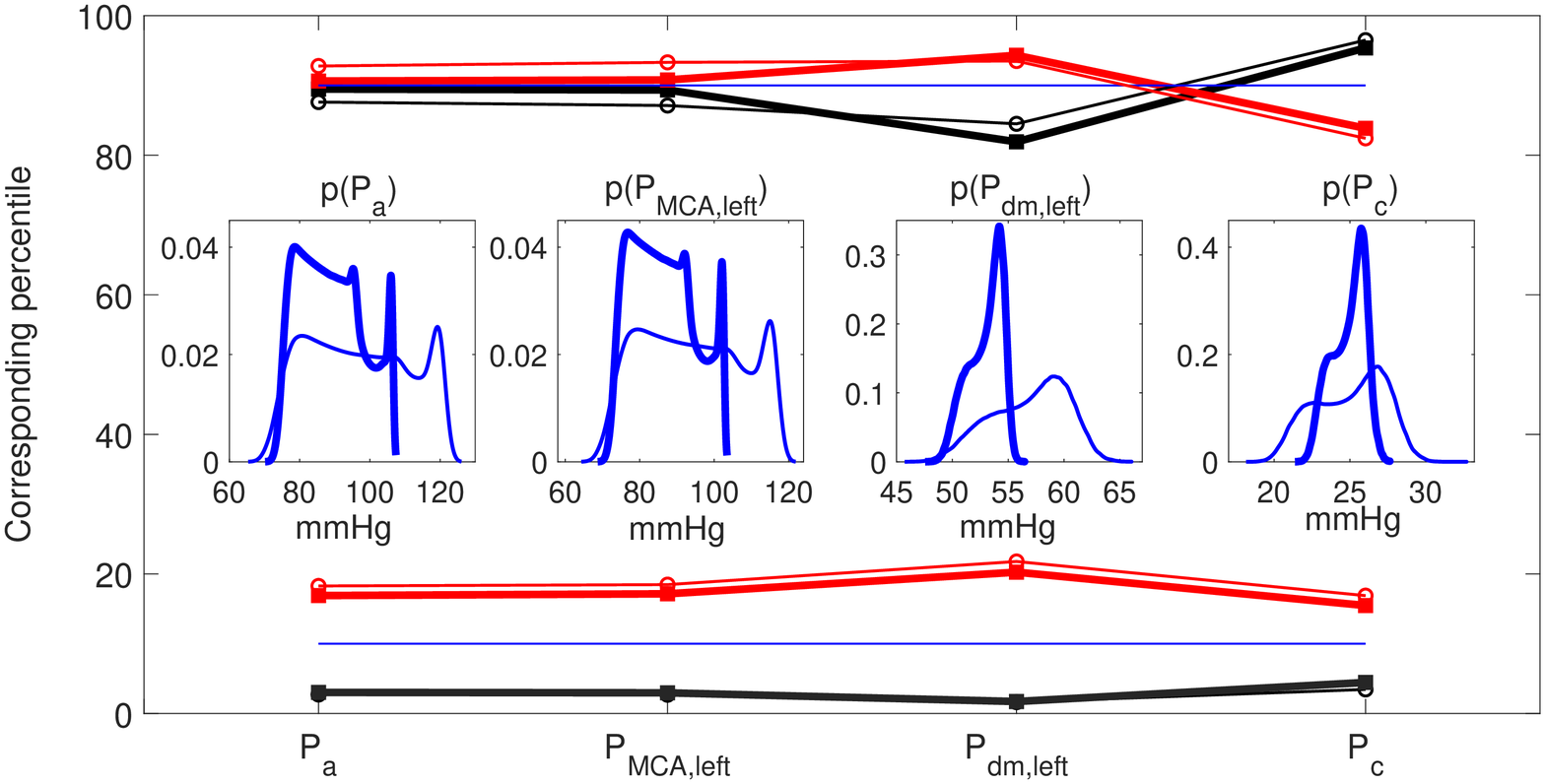}
\includegraphics[width=\columnwidth]{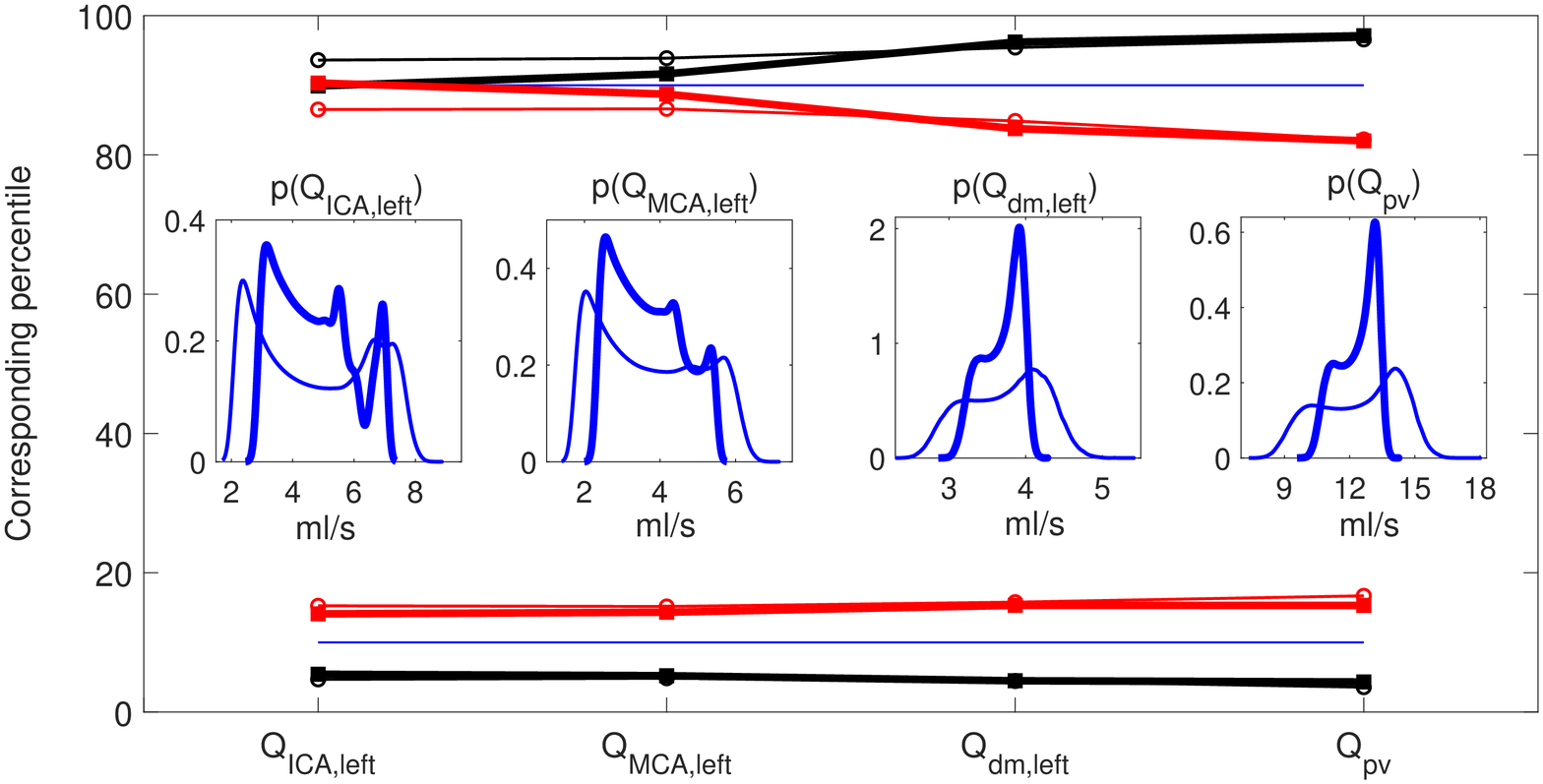}
\end{center}
\caption{Heart failure (thick) vs healthy (thin) percentile analyses. Percentile values in increased (red) and decreased (black) HRV to which the 10th and 90th percentiles in baseline configuration correspond. Pressure (top panel) and flow rate (bottom panel) along the ICA-MCA pathway. Inset panels represent the PDFs of the continuous hemodynamic time-series.}
\label{Fig_perc_HF}
\end{figure}

\begin{table}
\centering
\caption{Basic statistics (mean: $\mu$, standard deviation: $\sigma$, coefficient of variation: $c_v$) of the continuous time-series along the ICA-MCA pathway: $P_a$, $Q_{ICA,left}$, $P_{MCA,left}$, $Q_{MCA,left}$, $P_{dm,left}$, $Q_{dm,left}$, $P_{c}$, and $Q_{pv}$.}
\bigskip
\begin{tabular}{l ccc ccc}
\toprule
\textbf{Configuration} & \multicolumn{3}{c}{\textbf{Pressure P [mmHg]}} & \multicolumn{3}{c}{\textbf{Flow rate Q [ml/s]}} \\
\cmidrule(lr){2-4} \cmidrule(lr){5-7}
                       & \textbf{$\mu$} & \textbf{$\sigma$}  & \textbf{$c_v$} & \textbf{$\mu$} & \textbf{$\sigma$}  & \textbf{$c_v$}  \\
\midrule
 & \multicolumn{3}{c}{\emph{$P_a$}} & \multicolumn{3}{c}{\emph{$Q_{ICA,left}$}} \\
\midrule
Decreased HRV (-20\%)                     & 99.30 & 13.40 & 0.13    &    4.74 & 1.68 & 0.35 \\
Baseline                 & 96.91 & 14.42 & 0.15    &    4.73 & 1.83 & 0.39 \\
Increased HRV (+20\%)                     & 94.41 & 15.40 & 0.16    &    4.72 & 1.99 & 0.42 \\
\midrule
 & \multicolumn{3}{c}{\emph{$P_{MCA,left}$}} & \multicolumn{3}{c}{\emph{$Q_{MCA,left}$}} \\
\midrule
Decreased HRV (-20\%)                     & 96.60 & 12.51 & 0.13    &    3.75 & 1.22 & 0.33 \\
Baseline                 & 94.22 & 13.44 & 0.14    &    3.74 & 1.35 & 0.36 \\
Increased HRV (+20\%)                     & 91.72 & 14.32 & 0.16    &    3.72 & 1.48 & 0.40 \\
\midrule
 & \multicolumn{3}{c}{\emph{$P_{dm,left}$}} & \multicolumn{3}{c}{\emph{$Q_{dm,left}$}} \\
\midrule
Decreased HRV (-20\%)                     & 58.11 & 2.79  & 0.05    &    3.75 & 0.43 & 0.11 \\
Baseline                 & 56.91 & 3.38  & 0.06    &    3.74 & 0.51 & 0.14 \\
Increased HRV (+20\%)                     & 55.66 & 3.97  & 0.07    &    3.72 & 0.60 & 0.16 \\
\midrule
 & \multicolumn{3}{c}{\emph{$P_{c}$}} & \multicolumn{3}{c}{\emph{$Q_{pv}$}} \\
\midrule
Decreased HRV (-20\%)                     & 25.02 & 1.85  & 0.07    &   12.49 & 1.52 & 0.12 \\
Baseline                 & 25.00 & 2.33  & 0.09    &   12.46 & 1.87 & 0.15 \\
Increased HRV (+20\%)                     & 24.97 & 2.85  & 0.11    &   12.43 & 2.24 & 0.18 \\
\bottomrule
\end{tabular}
\label{complete_statistics}
\end{table}

\begin{table}
\centering
\caption{Mean and standard deviation (in brackets) values for the minimum ($P^{min}$ and $Q^{min}$), average ($\overline{P}$ and $\overline{Q}$), maximum ($P^{max}$ and $Q^{max}$), and pulse ($P^{pv}$ and $Q^{pv}$) values per beat of $P$ and $Q$ along the ICA-MCA pathway: $P_a$, $Q_{ICA,left}$, $P_{MCA,left}$, $Q_{MCA,left}$, $P_{dm,left}$, $Q_{dm,left}$, $P_{c}$, and $Q_{pv}$.}
\bigskip
\begin{tabular}{l cc cc}
\toprule
 \textbf{Configuration} & \textbf{Min value} & \textbf{Mean value}  & \textbf{Max value} & \textbf{Pulse value} \\
\midrule
  & $P_a^{min}$ & $\overline{P}_a$ & $P_a^{max}$ & $P_a^{pv}$ \\
\midrule
Decreased HRV (-20\%)                     & 78.43 (2.65) & 99.42 (1.86) & 121.26 (1.36) & 42.82 (1.56) \\
Baseline                 & 74.52 (3.02) & 97.06 (2.19) & 120.45 (1.64) & 45.93 (1.72) \\
Increased HRV (+20\%)                     & 70.63 (3.31) & 94.59 (2.47) & 119.54 (1.82) & 48.91 (1.89) \\
\midrule
  & $P_{MCA,left}^{min}$ & $\overline{P}_{MCA,left}$ & $P_{MCA,left}^{max}$ & $P_{MCA,left}^{pv}$ \\
\midrule
Decreased HRV (-20\%)                     & 77.08 (2.58) & 96.71 (1.83) & 117.11 (1.36) & 40.03 (1.44) \\
Baseline                 & 73.27 (2.94) & 94.36 (2.16) & 116.16 (1.63) & 42.89 (1.58) \\
Increased HRV (+20\%)                     & 69.48 (3.24) & 91.90 (2.43) & 115.11 (1.81) & 45.63 (1.73) \\
\midrule
  & $P_{dm,left}^{min}$ & $\overline{P}_{dm,left}$ & $P_{dm,left}^{max}$ & $P_{dm,left}^{pv}$ \\
\midrule
Decreased HRV (-20\%)                     & 53.72 (1.56) & 58.16 (1.30) & 61.04 (1.25) & 7.33 (0.82) \\
Baseline                 & 51.59 (1.84) & 56.97 (1.52) & 60.54 (1.50) & 8.95 (1.02) \\
Increased HRV (+20\%)                     & 49.34 (2.05) & 55.73 (1.65) & 60.05 (1.61) &	10.71 (1.21) \\
\midrule
  & $P_{c}^{min}$ & $\overline{P}_{c}$ & $P_{c}^{max}$ & $P_{c}^{pv}$ \\
\midrule
Decreased HRV (-20\%)                     & 22.05 (0.74) & 25.03 (0.66) & 27.12 (0.78) &	5.07 (0.64) \\
Baseline                 & 21.29 (0.90) & 25.01 (0.77) & 27.70 (0.93) &	6.41 (0.81) \\
Increased HRV (+20\%)                     & 20.45 (0.99) & 24.98 (0.83) & 28.37 (1.07) &	7.92 (0.99) \\
\midrule
  & $Q_{ICA,left}^{min}$ & $\overline{Q}_{ICA,left}$ & $Q_{ICA,left}^{max}$ & $Q_{ICA,left}^{pv}$ \\
\midrule
Decreased HRV (-20\%)                     & 2.32 (0.16) & 4.75 (0.14) & 7.30 (0.24) & 4.99 (0.23) \\
Baseline                 & 2.14 (0.15) & 4.74 (0.17) & 7.53 (0.28) & 5.40 (0.29) \\
Increased HRV (+20\%)                     & 1.98 (0.15) & 4.73 (0.18) & 7.80 (0.33) & 5.81 (0.34) \\
\midrule
  & $Q_{MCA,left}^{min}$ & $\overline{Q}_{MCA,left}$ & $Q_{MCA,left}^{max}$ & $Q_{MCA,left}^{pv}$ \\
\midrule
Decreased HRV (-20\%)                     & 1.96 (0.13) & 3.75 (0.13) & 5.76 (0.21) & 3.81 (0.20) \\
Baseline                 & 1.81 (0.14) & 3.74 (0.15) & 5.96 (0.26) & 4.16 (0.25) \\
Increased HRV (+20\%)                     & 1.66 (0.14) & 3.73 (0.16) & 6.17 (0.29) & 4.51 (0.29) \\
\midrule
  & $Q_{dm,left}^{min}$ & $\overline{Q}_{dm,left}$ & $Q_{dm,left}^{max}$ & $Q_{dm,left}^{pv}$ \\
\midrule
Decreased HRV (-20\%)                     & 3.06 (0.17) & 3.75 (0.15) & 4.23 (0.18) & 1.17 (0.13) \\
Baseline                 & 2.91 (0.19) & 3.74 (0.18) & 4.33 (0.22) & 1.42 (0.16) \\
Increased HRV (+20\%)                     & 2.75 (0.20) & 3.73 (0.19) & 4.44 (0.24) & 1.68 (0.19) \\
\midrule
  & $Q_{pv}^{min}$ & $\overline{Q}_{pv}$ & $Q_{pv}^{max}$ & $Q_{pv}^{pv}$ \\
\midrule
Decreased HRV (-20\%)                     & 9.98 (0.52) & 12.49 (0.43) &	14.26 (0.53) & 4.28 (0.46) \\
Baseline                 & 9.41 (0.62) & 12.47 (0.48) &	14.70 (0.60) & 5.29 (0.56) \\
Increased HRV (+20\%)                     & 8.82 (0.66) & 12.44 (0.51) &	15.19 (0.68) & 6.37 (0.66) \\
\bottomrule
\end{tabular}
\label{beat_to_beat}
\end{table}

\begin{table}
\centering
\caption{Total number of one-beat extreme events (out of 5000 RR beats) along the ICA-MCA pathway for the two decreased and increased HRV configurations.}
\bigskip
\begin{tabular}{l cccc cccc}
\toprule
\textbf{Configuration} & \multicolumn{4}{c}{\textbf{Pressure $\overline{P}$}} & \multicolumn{4}{c}{\textbf{Flow rate $\overline{Q}$}} \\
 \cmidrule(lr){2-5} \cmidrule(lr){6-9}
                       & \textbf{$\overline{P}_a$} & \textbf{$\overline{P}_{MCA,left}$}  & \textbf{$\overline{P}_{dm,left}$} & \textbf{$\overline{P}_c$} & \textbf{$\overline{Q}_{ICA,left}$}  & \textbf{$\overline{Q}_{MCA,left}$} & \textbf{$\overline{Q}_{dm,left}$} & \textbf{$\overline{Q}_{pv}$}  \\
\midrule
& \multicolumn{4}{c}{Hypotensive events} & \multicolumn{4}{c}{Hypoperfusion events}\\
\midrule
Decreased HRV (-20\%)                   & 0 & 0 & 0 & 0    &    0 & 0 & 0 & 0 \\
Increased HRV (+20\%)                   & 0 & 0 & 42 & 0   &    0 & 0 & 0 & 0 \\
\midrule
& \multicolumn{4}{c}{Hypertensive events} & \multicolumn{4}{c}{Hyperperfusion events}\\
\midrule
Decreased HRV (-20\%)                   & 0 & 0 & 69 & 0    &    0 & 0 & 0 & 0 \\
Increased HRV (+20\%)                   & 0 & 0 & 4 & 6     &    0 & 0 & 9 & 0 \\
\bottomrule
\end{tabular}
\label{hypo_hyper}
\end{table}

\begin{table}
\centering
\caption{Cardiac parameters and oxygen consumption indexes at central level: mean (std). $V_{lved}$: end-diastolic left ventricular volume, $V_{lves}$: end-systolic left ventricular volume, $SV$: stroke volume, $EF$: ejection fraction, $CO$: cardiac output, $SW/min$: stroke work per minute, $RPP$: rate pressure product, $TTI/min$: tension time index per minute, $PVA/min$: pressure volume area per minute, $LVE$: left ventricle efficiency.}
\bigskip
\begin{tabular}{l ccc}
\toprule
\textbf{Variable} & \textbf{Decreased HRV (-20\%)} & \textbf{Baseline} & \textbf{Increased HRV (+20\%)}\\
\midrule
$V_{lved}$ [ml]        & 126.99 (1.71) & 130.87 (1.79) & 134.43 (1.95) \\
\midrule
$V_{lves}$ [ml]        & 58.80 (1.16)  & 56.73 (1.30)  & 54.78 (1.38) \\
\midrule
$SV$ [ml]           & 68.18 (2.77)  & 74.14 (2.97)  & 79.65 (3.22) \\
\midrule
$EF$ [\%]           & 53.67 (1.49)  & 56.63 (1.53)  & 59.23 (1.58) \\
\midrule
$CO$ [l/min]        & 5.52 (0.30)   & 5.21 (0.32)   & 4.91 (0.31)  \\
\midrule
$SW/min$ [J/min]    & 75.44 (4.84)  & 69.46 (5.01)  & 63.72 (4.98) \\
\midrule
$RPP$ [mmHg/min]    & 9837.29 (835.45) & 8494.19 (786.38) & 7393.16 (733.68) \\
\midrule
$TTI/min$ [mmHgs/min] & 2775.53 (100.87) & 2622.53 (102.08) & 2488.01 (101.07) \\
\midrule
$PVA/min$ [J/min]      & 101.85 (7.15) & 91.98 (7.26) & 82.93 (7.13) \\
\midrule
$LVE$                & 0.74 (0.01) & 0.76 (0.01) & 0.77 (0.01) \\
\bottomrule
\end{tabular}
\label{central_statistics}
\end{table}

\begin{table}
\centering
\caption{Heart failure. Cardiac parameters and oxygen consumption indexes at central level: mean (std). $V_{lved}$: end-diastolic left ventricular volume, $V_{lves}$: end-systolic left ventricular volume, $SV$: stroke volume, $EF$: ejection fraction, $CO$: cardiac output, $SW/min$: stroke work per minute, $RPP$: rate pressure product, $TTI/min$: tension time index per minute, $PVA/min$: pressure volume area per minute, $LVE$: left ventricle efficiency.}
\bigskip
\begin{tabular}{l ccc}
\toprule
\textbf{Variable} & \textbf{Decreased HRV (-20\%)} & \textbf{Baseline} & \textbf{Increased HRV (+20\%)}\\
\midrule
$V_{lved}$ [ml]        & 251.07 (1.16) & 253.01 (1.11) & 254.51 (1.01) \\
\midrule
$V_{lves}$ [ml]        & 213.19 (0.81) & 211.47 (1.17) & 209.43 (1.56) \\
\midrule
$SV$ [ml]              & 37.88 (1.87) & 41.54 (2.19)  & 45.08 (2.47) \\
\midrule
$EF$ [\%]              & 15.08 (0.68)	& 16.41 (0.80)	& 17.71 (0.90)\\
\midrule
$CO$ [l/min]           & 3.64 (0.17) & 3.55 (0.17) & 3.46 (0.18) \\
\midrule
$SW/min$ [J/min]       & 39.22 (1.95) &	37.55 (1.97) & 35.98 (1.99) \\
\midrule
$RPP$ [mmHg/min]       & 10243.26 (699.45) & 9116.87 (691.62) &	8189.49 (672.01) \\
\midrule
$TTI/min$ [mmHgs/min]  & 2709.65 (55.37) & 2621.87 (58.59) &	2543.39 (61.17) \\
\midrule
$PVA/min$ [J/min]      & 132.13 (7.64) & 120.59 (7.81) & 110.73 (7.71) \\
\midrule
$LVE$                  & 0.30 (0.01) & 0.31 (0.01) & 0.33 (0.01) \\
\bottomrule
\end{tabular}
\label{central_statistics_HF}
\end{table}




\end{document}